\documentclass[aps,prd,preprintnumbers,superscriptaddress,nofootinbib,amsmath,amssymb,showpacs]{revtex4}
\usepackage{graphicx}
\usepackage{dcolumn}
\usepackage{bm}
\usepackage{amsmath}
\usepackage{color}

\DeclareMathAlphabet{\mathpzc}{OT1}{pzc}{m}{it}

\usepackage{amsfonts}

\begin{document}

\begin{flushright}
RESCEU-43/14
\end{flushright}
\title{Dynamics of cosmic strings with higher-dimensional windings}

\author{Daisuke Yamauchi}
\email[Email: ]{yamauchi``at''resceu.s.u-tokyo.ac.jp}
\affiliation{ Research Center for the Early Universe \\ School of Science, The University of Tokyo \\ 7-3-1 Hongo, Bunkyo-ku, Tokyo 113-0033, Japan}
\author{Matthew J. Lake}
\email[Email: ]{matthewj``at''nu.ac.th}
\affiliation{ The Institute for Fundamental Study, ``The Tah Poe Academia Institute", \\ Naresuan University, Phitsanulok 65000, Thailand}
\affiliation{ Thailand Center of Excellence in Physics, Ministry of Education, Bangkok 10400, Thailand}


\begin{abstract}
We consider F-strings with arbitrary configurations in the Minkowski directions of a higher-dimensional spacetime, 
which also wrap and spin around $S^1$ subcycles of constant radius in an arbitrary internal manifold, and 
determine the relation between the higher-dimensional and the effective four-dimensional
quantities that govern the string dynamics.
We show that, for any such configuration, the motion of the windings in the compact space may render the string 
effectively tensionless from a four-dimensional perspective, so that it remains static with respect to the large dimensions. 
Such a critical configuration occurs when (locally) exactly half the square of the string length lies in the large dimensions 
and half lies in the compact space. The critical solution is then seen to arise as a special case, in which the wavelength of the windings 
is equal to their circumference. As examples, long straight strings and circular loops are considered in detail, and the solutions 
to the equations of motion that satisfy the tensionless condition are presented. 
These solutions are then generalized to planar loops and arbitrary three-dimensional configurations. 
Under the process of dimensional reduction, in which higher-dimensional motion is equivalent 
to an effective worldsheet current (giving rise to a conserved charge), this phenomenon may be seen 
as the analogue of the tensionless condition which arises for superconducting and chiral-current carrying cosmic strings. 
\end{abstract}

\maketitle

%
\section{Introduction} \label{Sect1}
In Ref.~\cite{Ni79}, Nielsen demonstrated the equivalence of Nambu-Goto strings \cite{Go71,Na77} embedded in an $M^4 \times S^1$ bulk space and superconducting strings embedded in $M^4$ via dimensional reduction. The connections between dimensionally reduced strings and chiral strings, and between (non-chiral) superconducting strings and strings with chiral currents, were investigated in \cite{NiOl87} and \cite{BlOlVi01}, respectively. In Ref.~\cite{CoHiTu87}, Copeland, Hindmarsh and Turok showed that static solutions always exist for superconducting strings (with zero net charge) of arbitrary configuration. These occur when the effective tension vanishes locally at each point on the string. It is therefore implicit, within the above stated results, that ÒstaticÓ solutions should also exist for strings compactified on $M^4 \times S^1$. That is, critical solutions should exist in which the string configuration remains static with respect to the infinite directions, even if it continues to move in the compact space. In fact, it is precisely this movement (that, under dimensional reduction, is analogous to the flow of current in a superconducting, or chiral string), that causes the effective pressure/tension in the Minkowski directions to vanish.\\ 
\indent
We here present such a critical analysis for strings with arbitrary configurations in Minkowski space. A further advantage of this work is that the results should also hold for strings wrapping $S^1$ subcycles of constant radius in \emph{any} compactified geometry. The only caveat required is that the solutions must be dynamically stabilized, where necessary, if topological stability is not guaranteed (as it is in the case of genuine $M^4 \times S^1$ compactification). This is the case with the solutions obtained in \cite{LaWa10,LaYo12}, which are valid for strings wrapping great circles in the $S^3$ manifold which regularizes the conifold tip in the Klebanov-Strassler geometry \cite{KlSt00}, as well as for those wrapping cycles on a genuine $S^1$ internal space. \\
\indent
At present, if we wish to determine the conditions under which a general Nambu-Goto string with an embedding in $M^4 \times S^1$ becomes static with respect to the Minkowski directions, using results in the existing literature, we would first need to calculate the effective $(3+1)$-dimensional superconducting string action to which it corresponds. This involves rewriting the full $(4+1)$-dimensional Nambu-Goto Lagrangian as the product of a $(3+1)$-dimensional Nambu-Goto term
and a factor containing all terms in the higher-dimensional coordinate. Interpreting this as a correction incorporating the effects of a worldsheet current, the tensionless condition reduces to a constraint on the the latter.
This must then be translated back into an equivalent condition on the motion of the string in the higher-dimensional space.\\
\indent
Alternatively, we could take the approach given in \cite{Ni79}. 
The tensionless condition then follows from the general constraint equations as a specific case, though this possibility was not explicitly considered in the analysis presented therein. This is certainly far simpler than the first method, but also suffers from several drawbacks. Firstly, the standard gauge adopted in  \cite{Ni79} was chosen so as to simplify the form of the equations of motion as much as possible. While this is extremely useful (and elegant) for illustrating general results, such as the equivalence of $M^4 \times S^1$ embedded Nambu strings and superconducting strings in $M^4$, for which it was originally used in that paper, it is rather difficult to write down ansatzes for specific embeddings corresponding to simple string configurations. Exceptions are the circular string loop and infinite straight string, whose embeddings are manifestly obvious in most gauges. However, if we wish to write down ansatzes corresponding to, say, an elliptical configuration \cite{LiCh96}, or figure of eight string \cite{And02} in $M^4$ (not to mention any higher-order generalization), it is far simpler to use regular Cartesian or polar coordinates which remain fixed with respect to the background space. The same is true for a host of other exact solutions for Nambu-Goto strings in Minkowski and FLRW spacetimes \cite{And02,ViSh00}.\\
\indent
However, we believe it is reasonable to expect that any shape-maintaining, periodic, solution to the equations of motion for a string in $M^4$ can be generalized to the higher-dimensional case in $M^4 \times S^1$. That is, if a string maintains a certain configuration, be it a circle, an ellipse, a figure of eight, etc, in Minowksi space, adding higher-dimensional motion in the $S^1$ should not alter its $(3+1)$-dimensional shape, at least under certain conditions, in which this motion respects the existing symmetry of the string. If true, this implies that all known self-consistent ansatzes which give rise to exact (or numerical) solutions in $M^4$ have counterparts in $M^4 \times S^1$, to which the latter reduce when the radius of the compact space shrinks to zero. Furthermore, in light of previous results \cite{Ni79,CoHiTu87}, this strongly suggests that the tensionless condition will arise as a specific, critical, solution in each case.\\
\indent
It is therefore worthwhile to search for an analogue of the tensionless condition given explicitly in  \cite{CoHiTu87} (and which is present implicitly in the results given in \cite{Ni79,NiOl87,BlOlVi01}) without the need to adopt an effective action for strings compactified on $M^4 \times S^1$, or to impose an unduly complicated gauge condition, in which self-consistent ansatzes and physically intuitive exact solutions can be difficult to find. As such, we aim to develop a general treatment for $M^4 \times S^1$-compactified strings in which the existence of a universal tensionless condition, for strings with arbitrary configurations in the $M^4$ submanifold, is simple to demonstrate and in which existing, known solutions in $M^4$ can be readily generalized to the higher-dimensional case.\\
\indent
To this end, we adopt the static gauge throughout the following work, which identifies the time-like worldsheet coordinate, $\tau$, with the proper time, $t$. 
However, when determining the counterpart solution, in $M^4\times S^1$, of a general embedding in $M^4$, we do not adopt any additional gauge conditions. 
Although the resulting equations of motion can be quite complicated, one major advantage is that we may substitute any ansatz corresponding to a known, 
self-consistent solution in $M^4$ and determine the corresponding embedding in $S^1$ that allows the string to maintain its overall shape in $(3+1)$ dimensions. 
In other words, we can easily determine the form of the angular coordinate embedding required to maintain the consistency of the original ansatz, 
thus determining the higher-dimensional generalization of the known $(3+1)$-dimensional result. \\
\indent
One additional advantage of analyzing rotating wound string configurations is that, while the resulting helical structure (which qualitatively resembles a corkscrew), generates momentum in both the compact and non-compact directions, these are separately conserved due to the independence of the relevant Killing vectors. Therefore, under dimensional reduction, we are able to interpret the former as a conserved charge and the latter as the momentum associated with the flow of charge along the string, from a four-dimensional perspective  \cite{Ni79,NiOl87}. The wound string analysis therefore automatically generalizes the results obtained previously for current-carrying, but charge neutral cosmic strings \cite{CoHiTu87,Test3,Test4,BlOlVi01}, to the charged case. This, in turn, automatically includes all vorton-type models, which correspond to specific (circular) charged string configurations, stabilized by the momentum generated by the charge flow \cite{Test1,Test2,Carter(1990),Carter&Martin(1993),Larsen(1993),Martins&Shellard(1998),Carter&Davis(2000),LaWa10,LaYo12,Vortons(2013)}.\\
\indent
As an immediate corollary to the above statements, it follows that any known solution to the equations of motion for a Nambu-Goto string automatically generalizes to a (qualitatively similar) solution to the equations of motion for a  charged, current-carrying string, described by the appropriately modified effective action. Furthermore, in the thin-width ``wire" approximation, the same holds true for vacuum solutions in field-theoretic models of cosmic strings (for example, the Nielsen-Olesen solution for the Abelian-Higgs string \cite{NO}), which likewise generalize to charged, current-carrying solutions with the same basic configuration, but different (quantitative) time evolution. These generalizations automatically admit the existence of tensionless states.\\
\indent
The structure of this paper is then as follows. In Sec.~\ref{sec:Background} we present the necessary background regarding the string equations of motion. 
Sec.~\ref{sec:Energy-momentum tensor in warped geometries} deals with the general expression for the four-dimensional energy-momentum tensor, the conserved quantities,
and the definitions of certain model parameters (such as $\omega$, the fraction of the total string length which lies in the Minkowski directions and 
$r$, the fraction of the perpendicular velocity in the Minkowski directions), which are used throughout the remainder of the text.
In Sec.~\ref{sec:Dynamics in warped geometries}, we show the equivalence of the Euler-Lagrange equations for the extra dimension and
a far simpler assumption, namely $r=\omega$. 
We also demonstrate that a tensionless string with arbitrary shape \emph{always} exists for $\omega^2=r^2=1/2$.
In Sec.~\ref{sec:Long, straight strings} we consider long, straight strings with higher-dimensional windings as the simplest example system.
Circular and noncircular loops with higher-dimensional windings are considered in Secs.~\ref{sec:Circular loops} and \ref{sec:Non-circular loops}, respectively, and the general tensionless solution is given in Sec. \ref{The general tensionless state}.
A brief summary of the conclusions, and suggestions for further work are given in Sec.~\ref{sec:Conclusion}.

\section{Background}
\label{sec:Background}
%
We use the metric signature $(+----)$ and consider a background metric which gives rise to a line-element of the form
\begin{eqnarray} \label{Met_2.1}
ds^2 = g_{IJ} dx^{I}dx^{J} =\widetilde g_{\mu\nu}dx^{\mu}dx^{\nu} - R^2d\phi^2 =a^2dt^2-\Gamma_{ij}dx^idx^j.
\end{eqnarray}
where $I,J\in\left\{0,1,2,3,4\right\}$, $\mu,\nu \in \left\{0,1,2,3\right\}$, $i,j\in\{1,2,3,4\}$, 
$\phi \in [0,2\pi)$ is an angular coordinate in the internal space and $R$ is a (constant) radius.  The phenomenological ``warp factor", $a \in (0,1]$, takes account of the fact that the internal dimensions may be flux-compactified (as expected in string theory). In this case, $a^2<1$ represents the back-reaction of the fluxes on the large dimensions.
Here, $\widetilde g_{\mu\nu}$ is the four-dimensional metric, assumed to be conformally related to the Minkowski spacetime via
\begin{eqnarray}
&&\widetilde g_{\mu\nu}dx^\mu dx^\nu =a^2\eta_{\mu\nu}dx^\mu dx^\nu =a^2dt^2-h_{mn}dx^mdx^n.
\end{eqnarray}
with $m,n\in\left\{1,2,3\right\}$.
Note that we do not literally assume an $M^4 \times S^1$ type compactification. Rather, we consider an (almost) arbitrary compactification of a six-dimensional, space-like Calabi-Yau manifold, $(CY)^6$, the only caveat being that it contains at least one $S^1$ subcycle of constant radius, $R$. As any string windings necessarily wrap $1$-cycles, we consider the simplest scenario, in which the effective radius of the winding configuration remains constant. In this case, the part of the background metric ``seen" by the string is of the form given by eq.~\eqref{Met_2.1}, regardless of its macroscopic structure. \\
\indent 
In practice, the internal space may be far more complicated. For example, in \cite{LaWa10,LaYo12}, strings wrapping $S^1$ sub-cycles of constant radius at the tip of the Klebanov-Strassler geometry \cite{KlSt00} were considered. Here the target space manifold is $M^4 \times S^3$ and the line element is given by $ds^2 = a^2\eta_{\mu\nu}dx^{\mu}dx^{\nu} - R^2\Omega_3^2$, where $\Omega_3^2$ is the line-element on the unit three-sphere. In Hopf coordinates, this is given by $\Omega_3^2 = d\psi^2 + \sin^2(\psi)d\chi^2 + \cos^2(\psi)d\phi^2$ where $\psi \in [0,\pi)$ is the polar angle and $\chi,\phi \in [0,2\pi)$ are the two azimuthal angles. Taking  $\psi(\tau,\sigma)$, $\chi(\tau,\sigma)$ and $\phi(\tau,\sigma)$ as embedding coordinates for the string, it is clear that the value of $\psi$ controls the effective radius of the windings, which may, in principle, vary as a function of both $\tau$ and $\sigma$. For $\psi={\rm const}.$, the winding radius is also constant, and for $\psi=0$ in particular, it takes its maximum value, $R$, the radius of the $S^3$. Similar arguments hold true for more complicated manifolds, as long as they contain at least one $S^1$ submanifold of constant radius. The advantage of using an effective metric, representing the part of the background space ``seen" by the string, is that we do not need to make any further assumptions regarding the general structure of the internal space. Specifically, we do not assume that the string windings will (or will not) be topologically stabilized. \\
\indent
In the absence of additional fluxes, the basic string action is the Nambu-Goto action \cite{Go71,Na77} which, using a metric with negative signature, takes the form
\begin{eqnarray} \label{Act_2.1}
S = -\mathcal{T}\int d^2\zeta \sqrt{-\gamma}, 
\end{eqnarray}
where $\gamma$ is the the determinant of the induced metric on the world-sheet, 
\begin{eqnarray} \label{IndMet_2.1}
\gamma_{ab}(X) = g_{IJ}\left(X\right) \partial_{a} X^{I} \partial_{b} X^{J},
\end{eqnarray}
with $a,b \in \left\{0,1\right\}$, $\zeta^{0}=\tau,\zeta^{1}=\sigma$, which is equal to the world-sheet area and, from here on, we use $x^{I}$ to refer to spacetime coordinates and $X^{I}$ to refer to embedding coordinates.\\
\indent
Eq. (\ref{Act_2.1}) may be written in a canonical form by defining
\begin{eqnarray} \label{LangDens_2.1}
\mathcal{L} = -\mathcal{T}\sqrt{-\gamma},
\end{eqnarray}
as the Langrangian density, where
\begin{eqnarray} \label{FundTens_2.1}
\mathcal{T} = \frac{1}{2\pi \alpha'},
\end{eqnarray}
and $\alpha'$ is the Regge slope parameter, which is related to the (fundamental) string length-scale via
\begin{eqnarray} \label{StrLen_2.1}
l_{\rm st} = \sqrt{\alpha'},
\end{eqnarray}
in natural units ($G=1$, $c=1$, $\hbar=1$) which, hereafter, we adopt. Variation with respect to the embedding coordinates gives
\begin{eqnarray} \label{ActVar_2.1}
\delta S = \int_{\tau_i}^{\tau_f} d\tau \int_{\sigma_i}^{\sigma_f} d\sigma \left\{\frac{\partial}{\partial \zeta^a}
\left(\frac{\partial\mathcal{L}}{\partial (\partial_a X^I)}\delta X^I \right) - \left[\frac{\partial}{\partial \zeta^a}
\left(\frac{\partial\mathcal{L}}{\partial (\partial_a X^I)}\right) - \frac{\partial\mathcal{L}}{\partial X^I}\right]\delta X^I \right\} = 0,
\end{eqnarray}
which yields the canonical Euler-Lagrange equations, 
\begin{eqnarray} \label{CanEL_2.1}
\frac{\partial}{\partial \zeta^a}\mathcal{P}_I^a-\frac{\partial\mathcal{L}}{\partial X^I} = 0,
\label{eq:canonical EL eq}
\end{eqnarray}
with a boundary condition
\begin{eqnarray} \label{BoundTerm_2.1}
\Bigl[\mathcal{P}_I^\sigma\delta X^I\Bigr]_{\sigma_i}^{\sigma_f}  = 0,
\end{eqnarray} 
where we have introduced
\begin{eqnarray} 
\mathcal{P}_I^a=\frac{\partial\mathcal{L}}{\partial (\partial_a X^I)},
\end{eqnarray} 
to denote the canonical momentum of $X^I$ with respect to $\zeta^a$.
To satisfy the boundary term, we may impose 
Dirichlet, Neumann or periodic boundary conditions \cite{Zwi09}.
Alternatively, using the identity
$\delta(-\gamma) = (-\gamma)\gamma^{ab}\delta\gamma_{ab},$
the equation of motion above \eqref{eq:canonical EL eq} can be rewritten in the form \cite{And02},
\begin{eqnarray} \label{AltEL_2.1}
\frac{\partial}{\partial \zeta^{a}}\left(\sqrt{-\gamma}\gamma^{ab}g_{IJ}\partial_{b}X^{J}\right) 
-\frac{1}{2}\sqrt{-\gamma}\gamma^{cd}\frac{\partial g_{KL}}{\partial X^{I}}\partial_{c}X^{K}\partial_{d}X^{L}= 0.
\end{eqnarray}
Hereafter, we take $\sigma_i = 0$ and $\sigma_f=2\pi$ as the boundary values of $\sigma$ without loss of generality.

\section{Strings with higher-dimensional windings}
\label{sec:Strings with higher-dimensional windings}

In this section, we discuss general properties of strings with higher-dimensional
windings.
We begin by the defining the energy-momentum tensor
in the whole spacetime, and quantities associated with physical observables, 
in Sec. \ref{sec:Energy-momentum tensor in warped geometries}.
Then, in Sec. \ref{sec:Dynamics in warped geometries}, we derive
the equations of motion and show the equivalence of the dynamical equation
for the extra-dimension with the four-dimensional ones, together with the simple condition $r=\omega$.

\subsection{Energy-momentum tensor}
\label{sec:Energy-momentum tensor in warped geometries}

In general, the spacetime energy-momentum tensor is defined via
\begin{eqnarray} \label{GenEMT_2.2}
T^{IJ} = \frac{-2}{\sqrt{-g}}\frac{\delta S}{\delta g_{IJ}},
\end{eqnarray}
so that, for the Nambu-Goto action in $(4+1)$-dimensions \cite{And02,ViSh00},
\begin{eqnarray} \label{NGActEMT_2.2}
T^{IJ} = \frac{1}{\sqrt{-g}}\mathcal{T}\int \sqrt{-\gamma} \gamma^{ab}\partial_{a}X^{I}\partial_{b}X^{J} \delta^5(x-X)d\tau d\sigma.
\end{eqnarray}
In the static gauge, $X^{0}=\zeta \tau$, where $\zeta$ is constant with dimensions of the length, this is
\begin{eqnarray} \label{SatGauEMT_2.2}
T^{IJ} = \frac{1}{\sqrt{-g}}\zeta^{-1}\mathcal{T}\int_0^{2\pi} \sqrt{-\gamma} \gamma^{ab}\partial_{a}X^{I}\partial_{b}X^{J} \delta^4(x-X)d\sigma,
\end{eqnarray}
Let $\ell$ denote the length of string in the interval $[0,\sigma]$ at time $t$, \cite{Zwi09}: 
\begin{eqnarray} \label{dl_2.4}
d\ell^2 =  \Gamma_{ij}\partial_{\sigma}X^i\partial_{\sigma}X^jd\sigma^2.
\end{eqnarray}
Since the quantities $\partial X^i/\partial\ell$ behaves as the unit vector tangent to the string direction, we can define
$v_\perp^i$ to be the component of the velocity $\dot X^i$ in the direction perpendicular to the string in terms of $\ell$:
\begin{eqnarray}
v_\perp^i =\dot X^i-\left(\Gamma_{jk}\dot X^j\frac{\partial X^k}{\partial\ell}\right)\frac{\partial X^i}{\partial\ell},
\label{eq:v_perp^i def}
\end{eqnarray}
where a dot ( $\dot{}$ ) indicates a derivative with respect to $t$. Moreover, we have
\begin{eqnarray}
v_\perp^2 =\Gamma_{ij}v_\perp^i v_\perp^j =\Gamma_{ij}\dot X^i\dot X^j-\left(\Gamma_{ij}\dot X^i\frac{\partial X^j}{\partial\ell}\right)^2.
\end{eqnarray}
For strings in the static gauge in warped geometries, the Nambu-Goto action may be rewritten in terms of $\ell$ and $v_\perp^2$ \cite{Zwi09}, as
\begin{eqnarray} \label{ActUnWarp_2.4}
S = -\mathcal{T}a\int dt d\sigma \left(\frac{d\ell}{d\sigma}\right)\sqrt{1-\frac{v_{\perp}^2}{a^2}}.
\end{eqnarray}
With these variables, we can rewrite the energy-momentum tensor as
\begin{eqnarray}
&&\sqrt{-g}T^{00}=\mathcal{T}\int_0^{2\pi} d\sigma\epsilon\,\delta^4 (x-X)
,\label{eq:T^00}\\
&&\sqrt{-g}T^{i0}=\mathcal{T}\int_0^{2\pi} d\sigma\epsilon v_\perp^i\,\delta^4 (x-X)
,\label{eq:T^i0}\\
&&\sqrt{-g}T^{ij}=\mathcal{T}\int_0^{2\pi} d\sigma \Bigl\{\epsilon v_\perp^i v_\perp^j -\epsilon^{-1}\partial_\sigma X^i\partial_\sigma X^j\Bigr\}\delta^4 (x-X)
,\label{eq:T^ij}
\end{eqnarray}
where $\epsilon$ is the energy per unit coordinate length defined by
\begin{eqnarray} 
\epsilon =\zeta\sqrt{-\gamma}\gamma^{\tau\tau}=\frac{1}{\sqrt{a^2-v_\perp^2}}\frac{d\ell}{d\sigma}.
\label{eq:epsilon def}
\end{eqnarray}
Once we obtain the length $d\ell$, and the perpendicular velocity $v_\perp^i$, we can immediately calculate the components of the energy-momentum tensor
using the expressions above.
Following Carter~\cite{Carter:1990nb}, we introduce the string tension $T$ and the string mass-energy per unit length $U$ by
\begin{eqnarray} 
\sqrt{-g}T^{IJ}=\int d\tau d\sigma\sqrt{-\gamma}\left( Uu^I u^J -Tn^I n^J\right)\delta^5 (x-X)
\,,\label{eq:Carter1}
\end{eqnarray}
where $g_{IJ}u^Iu^J=-g_{IJ}n^In^J=1$\,, $g_{IJ}u^In^J=0$\,. One can see that,
in the case of eqs.~\eqref{eq:T^00}-\eqref{eq:T^ij}, the vectors
\begin{eqnarray} 
u^I=
\frac{1}{\sqrt{a^2-v_\perp^2}}\left( 1,v_\perp^i\right)
\,,\ \ \ \ 
n^I=
\left(\frac{d\ell}{d\sigma}\right)^{-1}
\left( 0,\partial_\sigma X^i\right).
\label{eq:u n def}
\end{eqnarray}
satisfy the necessary conditions.
Comparing eqs.~\eqref{eq:T^00}-\eqref{eq:T^ij} and eq.~\eqref{eq:Carter1}, we have an ordinary equation of state,
namely $U=T=\mathcal{T}$\,.
Therefore, as far as we consider the full energy-momentum tensor, the tension
takes the same value as the mass-energy per unit length, as usual.
However, what we actually observe is the effective four-dimensional energy-momentum tensor, which is defined by
\begin{eqnarray}
\widetilde T^{\mu\nu}\equiv\int  T^{\mu\nu}\sqrt{g_{\phi\phi}}d\phi.
\label{eq:effective T^munu}
\end{eqnarray}
An important observation is that the velocity $u^\mu$ is no longer perpendicular to the string direction $n^\mu$ 
from the four-dimensional point of view, namely $\widetilde g_{\mu\nu}u^\mu n^\nu\neq 0$.
Hence, an orthogonal basis different from $\{u^\mu ,n^\mu\}$ should be chosen 
to define the four-dimensional effective tension and mass-energy.
This fact implies that the effective equation of state may be different from the ordinary one.
In what follows, the quantities defined on the four-dimensional spacetime will be distinguished from their counterparts on the full five-dimensional spacetime by
the presence of a tilde ( $\widetilde{}$ ), as above. An exception is the effective world-sheet current, as seen from a four-dimensional perspective, which we will label $j$. Since the true source of this current is the movement of the string in the extra dimension, it must be defined with respect to higher-dimensional variables, but we must remember that, under under dimensional reduction, it becomes an effective world-sheet current in $(3+1)$ dimensions.
\\
\indent
Since, even in an $M^4\times S^1$ geometry, we have a Killing vector along the time direction, the conserved charges can be considered.
Using the letters $\widetilde E$ and $\widetilde P$ to denote the four-dimensional energy and momentum, 
respectively, we therefore define these as
\begin{eqnarray} 
	&&\widetilde E\equiv a\int\sqrt{-\widetilde g}\,\widetilde T^{00}d^3x=\mathcal{T}a\int_0^{2\pi} d\sigma\epsilon
	,\\
	&&\widetilde P^m\equiv a\int\sqrt{-\widetilde g}\,\widetilde T^{m0}d^3x =\mathcal{T}a\int_0^{2\pi} d\sigma\epsilon v_\perp^m.
\end{eqnarray}
Note that we have another conserved charge associated with the motion along the compact space, which is given by 
\begin{eqnarray} 
	Q\equiv R\int\sqrt{-g}T^{\phi 0}d^4x=\mathcal{T}\int_0^{2\pi}d\sigma Rv_\perp^\phi.
\end{eqnarray}
\\
\indent
Let us introduce a new parameter to represent the fraction of the string lying in the large spatial dimensions.
The length of the string $d\ell^2$ is split into the sum of a three-dimensional and an extra-dimensional part, where
\begin{eqnarray} \label{dlsum_2.4}
d\ell^2 = d\widetilde\ell^2 + R^2(\partial_\sigma\phi )^2d\sigma^2
=h_{mn}\partial_\sigma X^m\partial_\sigma X^n d\sigma^2 +R^2(\partial_\sigma\phi )^2d\sigma^2,
\end{eqnarray}
where $\widetilde\ell$ is the length of string as seen from a four-dimensional perspective.
We then define the parameter $\omega \in [0,1]$, which represents the fraction of the string lying in the large spatial dimensions, via
\begin{eqnarray} \label{omega_2.4}
\omega^{-2}\equiv\left(\frac{d\ell}{d\widetilde\ell}\right)^2
= \frac{h_{mn}\partial_{\sigma}X^m\partial_{\sigma}X^n +R^2(\partial_\sigma\phi )^2}{h_{mn}\partial_{\sigma}X^m\partial_{\sigma}X^n}.
\end{eqnarray}
In terms of the four-dimensional string length $\widetilde\ell$, it is useful to introduce 
the vector $\partial X^m/\partial\widetilde\ell$, which is normalized in the four-dimensional space.
We then define $\tilde v^m$ in the direction perpendicular to $\partial_\sigma X^m$ as
\begin{eqnarray}
\widetilde v_\perp^m =v_\perp^m -\left( h_{pq}v_\perp^p\frac{\partial X^q}{\partial\widetilde\ell}\right)\frac{\partial X^m}{\partial\widetilde\ell}
=\dot X^m-\left( h_{pq}\dot X^p\frac{\partial X^q}{\partial\widetilde\ell}\right)\frac{\partial X^m}{\partial\widetilde\ell}.
\label{eq:tilde v_perp^m def}
\end{eqnarray}
In terms of $\widetilde v_\perp^m$, we further define the parameter $r \in [0,1]$ which represents the fraction of the velocity field lying in the large dimensions, and which is defined as
\begin{eqnarray}
r^{-2}\equiv\frac{a^2-\widetilde v_\perp^2}{a^2-v_\perp^2}=\frac{a^2-\widetilde v_\perp^2}{a^2-\widetilde v_\perp^2 -\frac{R^2}{\omega^2}(v_\perp^\phi )^2},
\label{eq: r def}
\end{eqnarray}
with $\widetilde v_\perp^2 =h_{mn}\widetilde v_\perp^m\widetilde v_\perp^n$.
In general, we may consider only two variables, $\omega$ and $r$, which contain all the information relating to the embedding in the compact space.
With these variables, we can define the four-dimensional orthogonal basis $\{\widetilde u^\mu ,\widetilde n^\mu\}$\,,
which satisfies the conditions $\widetilde g_{\mu\nu}\widetilde u^\mu\widetilde u^\nu =-\widetilde g_{\mu\nu}\widetilde n^\mu\widetilde n^\nu =1$
and $\widetilde g_{\mu\nu}\widetilde u^\mu\widetilde n^\nu =0$. The explicit expressions are
\begin{eqnarray}
&&\widetilde u^\mu =ru^\mu +\frac{\sqrt{(1-\omega^2 )(1-r^2)}}{\omega}\, n^\mu
=\frac{1}{\sqrt{a^2-\widetilde v_\perp^2}}\left( 1,\widetilde v_\perp^m\right)
,\label{eq:tilde u}\\
&&\widetilde n^\mu =\frac{1}{\omega}n^\mu 
=\left(\frac{d\widetilde\ell}{d\sigma}\right)^{-1}\left( 0,\partial_\sigma X^m\right),
\label{eq:tilde n}
\end{eqnarray}
where we have used eq.~\eqref{eq:u n def}.
Substituting eqs.~\eqref{eq:tilde u}-\eqref{eq:tilde n} into eqs.~\eqref{eq:Carter1} and \eqref{eq:effective T^munu}, we then obtain 
the effective four-dimensional energy-momentum tensor
\begin{eqnarray}
\sqrt{-\widetilde g}\,\widetilde T^{\mu\nu}=\int d\tau d\sigma\sqrt{-\gamma}
\biggl[\,\widetilde U\,\widetilde u^\mu\widetilde u^\nu 
+2\widetilde\Sigma\,\widetilde u^{ (\mu}\widetilde n^{\nu )}
-\widetilde T\,\widetilde n^\mu\widetilde n^\nu\,
\biggr]\delta^4 (x-X)
,\label{eq:tilde T^munu}
\end{eqnarray}
where the effective energy-mass, shear and tension can be described as
\begin{eqnarray}
&&\widetilde U=\mathcal{T}r^{-2}
,\ \ \ 
\widetilde\Sigma =\mathcal{T}\sqrt{(1-\omega^2 )(1-r^2)}
\,,\ \ 
\widetilde T=\widetilde U\left(\omega^2 +r^2-1\right)
.
\end{eqnarray}
These results imply
$\widetilde U\widetilde T=\mathcal{T}^2(\omega^2 +r^2-1)/r^2$, which in general does not coincide with $\mathcal{T}^2$.
We further find that there is a nonvanishing effective shear term $\widetilde\Sigma$ due to the higher dimensional winding.
One can also see that, in the case without windings, namely $\omega\rightarrow 1,r\rightarrow 1$, 
the effective tension $\widetilde T$ approaches the effective mass-energy per unit length $\widetilde U$,
implying that its equation of state coincides with the ordinary one.
When the higher dimensional windings are taken into account, the equation of state is, in general, different from
the ordinary one.
In particular, the effective tension vanishes when
\begin{eqnarray}
\omega^2 +r^2-1=0.
\label{eq:tensionless condition}
\end{eqnarray}

\subsection{Dynamics}
\label{sec:Dynamics in warped geometries}

In this subsection, we will explicitly show that the dynamical equation of motion for the extra dimension
is equivalent to the four-dimensional equations, together with the simple condition $r=\omega$.
To show this, let us start with writing down the canonical variables
in terms of the perpendicular velocity $v_\perp^i=(v_\perp^m ,v_\perp^\phi )$ as
\begin{eqnarray}
	&&\mathcal{P}_0^\tau =-\mathcal{T}\epsilon,\ \ \ \ \mathcal{P}_0^\sigma =-\mathcal{T}\zeta j
	,\\
	&&\mathcal{P}_i^\tau =-\mathcal{T}\sqrt{-\gamma}\,\Gamma_{ij}\left(\gamma^{\tau\tau}\partial_\tau X^j+\gamma^{\tau\sigma}\partial_\sigma X^j\right) 
		=-\mathcal{T}\epsilon\,\Gamma_{ij} v_\perp^j
	,\\
	&&\mathcal{P}_i^\sigma =-\mathcal{T}\sqrt{-\gamma}\,\Gamma_{ij}\left(\gamma^{\sigma\tau}\partial_\tau X^j +\gamma^{\sigma\sigma}\partial_\sigma X^j\right)
		=-\mathcal{T}\zeta\,\Gamma_{ij}\left( j\,v_\perp^j -\frac{1}{\epsilon}\partial_\sigma X^j\right),
\end{eqnarray}
where $\epsilon$ has been defined in eq.~\eqref{eq:epsilon def} and 
\begin{eqnarray}
	j=\sqrt{-\gamma}\gamma^{\tau\sigma}=-\frac{\Gamma_{ij}\dot X^i\partial_\sigma X^j}{\left(\frac{d\ell}{d\sigma}\right)\sqrt{a^2-v_\perp^2}},
\end{eqnarray}
which may be interpreted as the effective current density on the string, from a four-dimensional perspective, after fixing the remaining gauge degree of freedom. In terms of $\epsilon$ and $j$, the perpendicular velocity can be rewritten as
\begin{eqnarray}
	v_\perp^i =\dot X^i+\frac{j}{\epsilon}\partial_\sigma X^i.
\end{eqnarray}
Hereafter, for convenience, we use the Cartesian coordinate system, namely $h_{mn}=\delta_{mn}$, without the loss of generality. 
We then rewrite the canonical Euler-Lagrange equations \eqref{eq:canonical EL eq} as 
\begin{eqnarray}
	&&\mathcal{E}_t\equiv\dot\epsilon +\partial_\sigma j=0
	,\label{eq:EOM 0 component}\\
	&&\mathcal{E}^m=\left(\epsilon v_\perp^m\right)^{\cdot}+\partial_\sigma\left( j\,v_\perp^m\right) -\partial_\sigma\left(\frac{1}{\epsilon}\partial_\sigma X^m\right) =0
	,\label{eq:EOM m component}\\
	&&\mathcal{E}_\phi =\left(\epsilon v_\perp^\phi\right)^{\cdot}+\partial_\sigma\left( j\,v_\perp^\phi\right) -\partial_\sigma\left(\frac{1}{\epsilon}\partial_\sigma\phi\right) =0
	.\label{eq:EOM phi component}
\end{eqnarray}
Equation \eqref{eq:EOM phi component} is the dynamical equation to solve, while the other two equations \eqref{eq:EOM 0 component} and \eqref{eq:EOM m component}
should be solved as the constraint equations for $\phi$, simultaneously.
Although these equations are very complicated, we will show that there is an alternative, simple, condition which is
equivalent to the original equation of motion in $\phi$ \eqref{eq:EOM phi component}.
Let us assume the condition given by
\begin{eqnarray}
	\frac{d\ell}{d\sigma}\sqrt{1-v_\perp^2}=\frac{d\widetilde\ell}{d\sigma}\sqrt{1-\widetilde v_\perp^2}
	\ \ \ \ \Leftrightarrow \ \ \ 
	r=\omega
	.\label{eq:condition}
\end{eqnarray}
which implies that the area of worldsheet seen in the five dimensions is equivalent to one seen in four dimensions.
Imposing this constraint implies that all higher-dimensional terms are removed from $(-\gamma)$ at the level of the equation of motion.
Using the properties of $\omega$ and $r$ (see eqs.~\eqref{omega_2.4} and \eqref{eq: r def}),
the condition \eqref{eq:condition} is equivalent to
\begin{eqnarray}
	\epsilon v_\perp^\phi =\epsilon\dot\phi +j\partial_\sigma\phi =\partial_\sigma\phi .
	\label{eq:condition2}
\end{eqnarray}
Substituting this into the equation of motion for the compactified dimension, 
namely eq.~\eqref{eq:EOM phi component}, 
one can easily see that eq.~\eqref{eq:EOM phi component} 
is trivially satisfied.
Therefore, eq.~\eqref{eq:condition} is genuinely
equivalent to the original equation of motion in $\phi$.
In fact, it is far simpler to work directly with eq.~\eqref{eq:condition} as the equation of motion in $\phi$.
We should emphasize that this result does not depend on the choice of the gauge in the sense that, 
throughout the derivations above, we have adopted only the static gauge and 
thus have one more gauge degree of freedom left to fix.
\\
\indent
To construct the equations of motion as a self-consistent set, we explicitly derive 
the remaining equations of motion, which naturally include the effect of the higher-dimensional windings, as the constraint equations.
For convenience, we introduce the following variables:
\begin{eqnarray}
	\widetilde\epsilon =\frac{1}{\sqrt{a^2-\widetilde v_\perp^2}}\frac{d\widetilde\ell}{d\sigma}
	,\ \ \ \ \ 
	\widetilde j=-\frac{h_{mn}\dot X^m\partial_\sigma X^n}{\left(\frac{d\widetilde\ell}{d\sigma}\right)\sqrt{a^2-\widetilde v_\perp^2}}
	,\label{eq:tilde epsilon and j}
\end{eqnarray}
where $\widetilde\epsilon$ represents the energy per unit length, as seen in the large dimensions, and $\widetilde j$ is a mathematically convenient definition, which depends on $j$ and $\omega^2$.
With these, and the four-dimensional perpendicular velocity $\widetilde v_\perp^m =\dot X^m+(\widetilde j/\widetilde\epsilon )\partial_\sigma X^m$,
it is useful to rewrite eq.~\eqref{eq:EOM m component} as 
\begin{eqnarray}
	\dot{\widetilde v}^m_\perp +\left( 2\frac{j}{\epsilon}-\frac{\widetilde j}{\widetilde\epsilon}\right)\partial_\sigma\dot X^m
	+\Biggl[
		\left(\frac{j}{\epsilon}-\frac{\widetilde j}{\widetilde\epsilon}\right)^{\cdot}+\frac{1}{2}\partial_\sigma\left(\frac{j^2-1}{\epsilon^2}\right)
	\Biggr]\partial_\sigma X^m
	+\frac{j^2-1}{\epsilon^2}\partial_\sigma^2 X^m
	=0.
\end{eqnarray}
We then decompose this into the two parts, with one part parallel and one part perpendicular to the direction of the string, giving
\begin{eqnarray}
	&&\mathcal{E}_\parallel\equiv h_{mn}\mathcal{E}^m\partial_\sigma X^m
	\notag\\
	&&\quad
		=\biggl[
			\left(\frac{j}{\epsilon}-\frac{\widetilde j}{\widetilde\epsilon}\right)^{\cdot}\widetilde\epsilon^2\left( a^2-\widetilde v_\perp^2\right)
		\biggr]
		-\frac{1}{2}\partial_\sigma\biggl[
			\left(\frac{j^2-1}{\epsilon^2}-\frac{\widetilde j^2-1}{\widetilde\epsilon^2}\right)\widetilde\epsilon^2\left( a^2-\widetilde v_\perp^2\right)
		\biggr] =0,
	\\
	&&\mathcal{E}_\perp\equiv h_{mn}\mathcal{E}^m\widetilde v_\perp^n
	\notag\\
	&&\quad
		=\frac{1}{\widetilde\epsilon}\left(\dot{\widetilde\epsilon}+\partial_\sigma\widetilde j\right)\left( a^2-\widetilde v_\perp^2\right)
			+\left(\frac{j}{\epsilon}-\frac{\widetilde j}{\widetilde\epsilon}\right)\partial_\sigma\left(\widetilde v_\perp^2\right)
	\notag\\
	&&\quad\quad
			-\frac{1}{2}\Biggl\{
				\left(\frac{j}{\epsilon}-\frac{\widetilde j}{\widetilde\epsilon}\right)^2 +\left(\frac{1}{\epsilon^2}-\frac{1}{\widetilde\epsilon^2}\right)	
			\Biggr\}
			\Biggl\{
				\Bigl[\widetilde\epsilon^2\left( a^2-\widetilde v_\perp^2\right)\Bigr]^{\cdot}
				+\frac{\widetilde\epsilon}{\widetilde j}\partial_\sigma\Bigl[\widetilde j^2\left( a^2-\widetilde v_\perp^2\right)\Bigr]
			\Biggr\},
\end{eqnarray}
respectively. We now impose the condition \eqref{eq:condition} and rewrite these equations in terms of the quantities $\omega$.
Using the properties:
\begin{eqnarray}
	\epsilon =\frac{\widetilde\epsilon}{\omega^2}
	,\ \ \ \ \ 
	j=\frac{\widetilde j-1+\omega^2}{\omega^2},
\end{eqnarray}
the consistent set of the equations of motion is obtained as
\begin{eqnarray}
	&&\mathcal{E}_\phi\ :\ r=\omega\ \ \ \ \Leftrightarrow \ \ \ 
	\widetilde\epsilon\,\dot\phi =\left( 1-\widetilde j\right)\partial_\sigma\phi
	\,,\label{eq:phi reduced EOM}\\
	&&\mathcal{E}_t
		=\left(\frac{\widetilde\epsilon}{\omega^2}\right)^{\cdot}
		+\partial_\sigma\left(\frac{\widetilde j-1}{\omega^2}\right)
		=0
	,\label{eq:t EOM r=omega}\\
	&&\mathcal{E}_\parallel
		=-\Bigl[\left( 1-\omega^2\right)\widetilde\epsilon\left( a^2-\widetilde v_\perp^2\right)\Bigr]^{\cdot}
			+\partial_\sigma\Bigl[\left( 1-\omega^2\right)\left(\widetilde j-1\right)\left( a^2-\widetilde v_\perp^2\right)\Bigr] 
		=0
	,\label{eq:v EOM1 r=omega}\\
	&&\widetilde\epsilon\mathcal{E}_\perp
		=\left(\dot{\widetilde\epsilon}+\partial_\sigma\widetilde j\right)\left( a^2-\widetilde v_\perp^2\right)
	\notag\\
	&&\quad\quad\quad
			-\left( 1-\omega^2\right)
				\biggl\{
					\frac{1}{\widetilde\epsilon}\Big[\widetilde\epsilon^2\left( a^2-\widetilde v_\perp^2\right)\Bigr]^{\cdot}
					+\frac{1}{\widetilde j}\partial_\sigma\Bigl[\widetilde j^2\left( a^2-\widetilde v_\perp^2\right)\Bigr]
					+\partial_\sigma\left(\widetilde v_\perp^2\right)
				\biggr\}
		=0.
	\label{eq:v EOM2 r=omega}
\end{eqnarray}
Once one imposes the gauge condition and the functional form of the embedding for the large dimensions, one can calculate
the four-dimensional quantities such as $\widetilde v_\perp^2$, $\widetilde\epsilon$ and $j$ through 
eqs.~\eqref{eq:tilde v_perp^m def} and \eqref{eq:tilde epsilon and j}.
Substituting these into eqs.~\eqref{eq:phi reduced EOM}-\eqref{eq:v EOM2 r=omega}, the self-consistent set
of equations for a string with higher-dimensional windings can be obtained.
This allows us, in principle, to take \emph{any} known self-consistent ansatz for a configuration in $M^4$ 
(expressed in Cartesian coordinates), and to work out its higher-dimensional counterpart in $M^4 \times S^1$.
That is, we are able to determine the precise form of $\phi$ for which the string configuration retains its overall shape 
during its dynamical evolution. Furthermore, what is true for the higher-dimensional extension of $(3+1)$-dimensional Nambu-Goto strings
should also be true for the superconducting and chiral-current carrying extensions of vacuum string solutions in field-theoretic models of cosmic strings.
\\
\indent
Based on the analysis presented here, let us now discuss the condition for the vanishing effective tension $\widetilde T$, 
eq.~\eqref{eq:tensionless condition}. 
As we have showed above, eq.~\eqref{eq:condition} is actually equivalent to the equation of motion for $\phi$.
Hence, imposing $r=\omega$ hereafter, the effective mass density, shear and tension appearing in eq.~\eqref{eq:tilde T^munu} are given by
\begin{eqnarray}
\widetilde U=\mathcal{T}\omega^{-2}
,\ \ \ \ 
\widetilde\Sigma =\mathcal{T}\left( 1-\omega^2\right)
,\ \ \ \ 
\widetilde T=\mathcal{T}\,\frac{2\omega^2 -1}{\omega^2},
\end{eqnarray}
respectively. Therefore, the effective tension vanishes when
\begin{eqnarray}
\omega^2 =\frac{1}{2},
\label{eq:omega^2=1/2}
\end{eqnarray}
or, equivalently, when exactly half the square of the string length lies in the large dimensions and half lies in the compact space.
When the tension vanishes, we have
\begin{eqnarray}
\widetilde U=2\mathcal{T}
,\ \ \ \ 
\widetilde\Sigma =\frac{1}{2}\mathcal{T}
,\ \ \ \ 
\widetilde\epsilon =\frac{1}{2}\epsilon
,\ \ \ \ 
\widetilde j=\frac{1}{2}\left( j+1\right)
.
\end{eqnarray}
Equation \eqref{eq:condition}, or equivalently eq.~\eqref{eq:condition2}, 
is also precisely the condition needed to ensure that $T^{\phi\phi}=0$, so that there is no effective pressure in either the $\phi$-direction, 
or the ``$R$-direction". 
Physically, this means that there is no effective pressure which can act to change the radius of the windings in the compact space. \\
\indent
For strings compactified on an genuine $M^4 \times S^1$ manifold, where the $S^1$ is of constant radius, $R$, 
it is of course meaningless to talk about the string moving in the ``$R$-direction", but the metric eq.~\eqref{Met_2.1} 
is also valid, as an effective metric, for embeddings in more complex internal spaces, where the string wraps cycles of constant radius 
in the compact dimensions. 
For example, in \cite{LaWa10,LaYo12} and \cite{BlIg05}, strings wrapping great circles of the $S^3$ internal manifold that regularizes 
the conifold at the tip of the Klebanov-Strassler geometry, \cite{KlSt00}, were considered. However, in such a geometry, 
the effective winding radius of the string may, in principle, be a function of $t$ and $\sigma$ and, since the windings are not topologically stabilized, 
they must be stabilized (if at all), dynamically.\\
\indent
Although we did not explicitly include a term proportional to $dR^2$ in the metric considered in this paper, or allow $R$ 
to be a function of the worldsheet coordinates in the embedding, we made such a restriction purely for simplicity and, 
in principle, the radius of the windings can change, in an arbitrary internal manifold, as a function of both space and time. 
Thus, nonzero $\dot{R}$ would be associated with nonzero $T^{\phi\phi}$. 
Seen in this way, the condition above and the equation of motion in $\phi$, respectively, result from, and 
are necessary conditions for, our initial assumption regarding the constancy of the winding radius.

\section{Applications}

In this section, we give several examples of the utility of the formulae
given in the previous section.
In subsection \ref{sec:Long, straight strings}, we first consider long,
straight strings as the simplest example, and show that there is 
a consistent solution for the higher-dimensional windings which leads to
the effective tensionless condition.
Then, in subsection \ref{sec:Circular loops}, we consider circular loops
to see the effect of the perpendicular velocity. 

\subsection{Long, straight strings}
\label{sec:Long, straight strings}

Let us adopt the Cartesian coordinates for the background spacetime and consider the embedding
\begin{eqnarray} \label{Emb_3.2}
	X^I=\left( t(\tau )=\zeta \tau, x=0, y=0, z(\sigma)=\frac{\Delta}{2\pi} \sigma, \phi(\tau ,\sigma) \right)
	,\label{eq:long, straight string embedding}
\end{eqnarray}
which represents long, straight strings with arbitrary higher-dimensional windings. 
We now adopt the gauge condition to identify the spacelike worldsheet coordinate with a fixed
spatial coordinate in the background space.
It is clear to see that this embedding also satisfies
the standard gauge condition in the four-dimensional point of view, namely $h_{mn}\dot X^m\partial_\sigma X^n=0$.
We first give a brief discussion about the four-dimensional quantities.
For the vanishing higher-dimensional winding, namely $\phi ={\rm const.}$, the length of the string and
the perpendicular velocity are given by
\begin{eqnarray}
	\frac{d\widetilde\ell}{d\sigma}=a\frac{\Delta}{2\pi}
	,\ \ \ \ \ 
	\widetilde v_\perp^m =0.
\end{eqnarray}
Hence, we obtain the four-dimensional orthogonal basis for the energy-momentum tensor
$\widetilde u^\mu =(1,0,0,0)\,,\,\widetilde n^\mu =(0,0,0,1)$,
as well as the conserved mass-density and current, given by
\begin{eqnarray}
	\widetilde\epsilon =\frac{1}{a}\frac{d\widetilde\ell}{d\sigma}=\frac{\Delta}{2\pi}
	,\ \ \ \ \ 
	j = -1, \ (\widetilde j=0),
\end{eqnarray}
respectively. As we already mentioned in Sec.~\ref{sec:Energy-momentum tensor in warped geometries},
the equation of state for the string without the winding is given by $\widetilde U=\widetilde T=\mathcal{T}$ 
and coincides with the ordinary one.
\\
\indent
Now let us consider the dynamics along the compact space, namely the higher-dimensional winding.
Since the four-dimensional embedding, $X^\mu$, always coincides with the solution of the 
embedding without the higher-dimension winding (in the sense that it represents the same basic shape, and hence has the same basic functional form), we have $\widetilde{\mathcal{E}}_\parallel =\widetilde{\mathcal{E}}_\perp =0$.
Taking the higher-dimensional winding into account with the condition $r=\omega$, 
we then have the length of the string and the perpendicular velocity as
\begin{eqnarray}
	\frac{d\ell}{d\sigma} =a\frac{\Delta}{2\pi}\omega^{-1}
	,\ \ \ \ \ 
	v_\perp^i =\left( 0,0,1-\omega^2,\frac{a\omega}{R}\sqrt{1-\omega^2}\right),
\end{eqnarray}
which leads to
\begin{eqnarray}
	\frac{v_\perp^2}{a^2} =\omega^2\dot\phi^2,
\end{eqnarray}
where, for this embedding, the functions $\omega$ and $r$ are given by
\begin{eqnarray} \label{omega_3.2}
	\omega =\left( 1+\frac{(2\pi )^2R^2}{a^2\Delta^2}(\partial_{\sigma}\phi)^2\right)^{-1/2}
	,\label{eq:long string omega}\ \ \ \ \ 
	r=\left( 1-\frac{R^2}{a^2}\omega^2\dot\phi^2\right)^{1/2}\,.
\end{eqnarray}
We now try to obtain the solution for the higher-dimensional winding that is consistent with the tensionless condition.
The relevant components of the constraint equations \eqref{eq:t EOM r=omega}-\eqref{eq:v EOM2 r=omega} are written as
\begin{eqnarray}
	\widetilde\epsilon\,\dot\omega =\partial_\sigma\omega =0.
\end{eqnarray}
Moreover, the condition \eqref{eq:condition} is reduced to
\begin{eqnarray}
	\widetilde\epsilon\,\dot\phi =\partial_\sigma\phi.
	\label{eq:long string r=omega}
\end{eqnarray}
Comparing eqs.~\eqref{eq:long string omega}-\eqref{eq:long string r=omega} and imposing
the periodic boundary condition on $\phi$, that is $\phi (t,\sigma +2\pi )=\phi (t,\sigma )$, 
we obtain the consistent solution
for the extra-dimension:
\begin{eqnarray}
	\phi =n_z\sigma +n_z\widetilde\epsilon^{-1} t\ \ \ \ \ n_z \in \mathbb{Z},
\end{eqnarray}
where $n_z\widetilde\epsilon^{-1}$ corresponds physically to an angular frequency for the rotation in the compact space, and 
where we have introduced $n_z \in \mathbb{Z}$, defined by
\begin{eqnarray} \label{n_z_3.2}
	n_z \equiv \frac{1}{2\pi}\int_{0}^{2\pi}\partial_{\sigma}\phi\, d\sigma,
\end{eqnarray}
to denote the net number of windings, which are distributed along the $z$-direction.
Substituting eq.~\eqref{n_z_3.2} into eq.~\eqref{eq:long string omega}, we have
\begin{eqnarray} 
	\omega^{-2}=r^{-2}=1+\left(\frac{2\pi Rn_z}{a\Delta}\right)^2 ={\rm const.}
\end{eqnarray}
It is straightforward to see that the effective four-dimensional mass-density, shear and tension are given by
\begin{eqnarray} 
	\widetilde U=\mathcal{T}\Biggl[ 1+\left(\frac{2\pi Rn_z}{a\Delta}\right)^2\Biggr]
	,\ \ 
	\widetilde\Sigma =\mathcal{T}\Biggl[ 1+\left(\frac{a\Delta}{2\pi Rn_z}\right)^2\Biggr]^{-1}
	,\ \ 
	\widetilde T=\mathcal{T}\Biggl[ 1-\left(\frac{2\pi Rn_z}{a\Delta}\right)^2\Biggr],
	\label{eq:U Sigma T in straight string}
\end{eqnarray}
respectively. Although the shear $\widetilde\Sigma$ appears in the effective energy-momentum tensor, 
the equation of state for this string seems to be the fixed trace type~\cite{Carter:1990nb}.
Clearly, it is possible for the string to be effectively tensionless, everywhere, when
the four-dimensional length of the string is equal to the circumference of the extra-dimension, that is
\begin{eqnarray} \label{tensionless1}
	a\Delta =2\pi Rn_z.
\end{eqnarray} 
Therefore, defining the wavelength of the windings with respect to the four-dimensional spacetime, $\lambda_z$, via
\begin{eqnarray} \label{lambda_z}
	\lambda_z = \frac{\Delta}{n_z},
\end{eqnarray} 
it is clear that the tensionless condition is equivalent to
\begin{eqnarray} \label{tensionless2}
	a\lambda_z =2\pi R.
\end{eqnarray} 
Furthermore, since $a \lambda_z$ (rather than simply $\lambda_z$) is the true effective wavelength with respect to a \emph{warped} 
background, we see that the tensionless case corresponds to a solution in which this is equal to the circumference of the windings, as claimed.

\subsection{Circular loops}
\label{sec:Circular loops}

In this subsection, we treat circular loops with higher-dimensional windings and give the general solution to the equation of motion. 
For circular loops with higher-dimensional windings, the embedding in the Cartesian coordinate system is
\begin{eqnarray} \label{Emb_4.1}
	X^I=\left( t(\tau)=\zeta \tau, x(\tau ,\sigma )=\rho (\tau )\cos\sigma ,y(\tau ,\sigma )=\rho (\tau )\sin\sigma ,z=0, \phi (\tau ,\sigma )\right).
\end{eqnarray}
One can easily see that the four-dimensional standard gauge $h_{mn}\dot X^m\partial_\sigma X^n=0$ can be applied to the above embedding.
Again, before proceeding to this analysis, it is useful to review the general solution of the equation of motion 
for circular loops without higher-dimensional windings.
In this case, the length of string and the perpendicular velocity given in Sec.~\ref{sec:Energy-momentum tensor in warped geometries} are
\begin{eqnarray} \label{dl_4.1}
	\frac{d\widetilde\ell}{d\sigma} = a\rho
	,\ \ \ \ \ 
	\widetilde v_\perp^m =\left(\dot\rho\cos\sigma ,\dot\rho\sin\sigma ,0\right).
\end{eqnarray}
We then obtain the energy per unit length,
\begin{eqnarray}
	\widetilde\epsilon =\frac{\rho}{\sqrt{1-\dot\rho^2}}.
\end{eqnarray}
When we consider the embedding without the higher-dimensional winding in a more general form, i.e. as $X^\mu$, 
the equations of motion are simply those for the harmonic oscillator.
It is then straightforward to show that, for a circular string, the equation of motion for the radial coordinate in the large dimensions reduces to the simple form:
\begin{eqnarray}
	\ddot\rho =-\widetilde\epsilon^{-2}\rho.
\end{eqnarray}
Since $\widetilde j=0$ in this embedding, $\widetilde\epsilon$ is treated as the conserved quantity. 
Hence, the general solution is the periodic function of the form
\begin{eqnarray}
	\rho\propto\Bigl|\cos\left(\widetilde\epsilon^{-1}t+(\text{phase})\right)\Bigl|,
\end{eqnarray}
where the amplitude and the phase of the solution are determined by the boundary conditions for $\rho$.
The loop performs one full oscillation (returning back to its original radius) with time period
$\widetilde\epsilon\pi$.
Although the energy-momentum tensor is not explicitly shown, the equation of state for the unwound string
is the ordinary one, as mentioned in Sec.~\ref{sec:Energy-momentum tensor in warped geometries}.

Let us now solve the full equations of motion including the higher-dimensional windings. 
In terms of $\omega =r$, the length and the perpendicular velocity are given by
\begin{eqnarray} \label{dl_4.2}
	&&\frac{d\ell}{d\sigma} =a\rho\omega^{-1}
	,\\
	&&v_\perp^i 
		=\biggl(
			\dot\rho\cos\sigma +\left( 1-\omega^2\right)\sqrt{1-\dot\rho^2}\sin\sigma ,
	\notag\\
	&&\quad
			\dot\rho\sin\sigma -\left( 1-\omega^2\right)\sqrt{1-\dot\rho^2}\cos\sigma ,
			0,\frac{a\omega}{R}\sqrt{\left( 1-\omega^2\right)\left( 1-\dot\rho^2\right)}
		\biggr)
\end{eqnarray}
so that
\begin{eqnarray}
	\frac{v_\perp^2}{a^2}=\dot\rho^2 +\omega^2\dot\phi^2.
\end{eqnarray}
Here, $\omega$ and $r$ for this embedding are given by
\begin{eqnarray}
	\omega =\left( 1+\frac{R^2}{a^2\rho^2}\left(\partial_\sigma\phi\right)^2\right)^{-1/2}
	,\ \ \ 
	r=\left( 1-\frac{R^2}{a^2(1-\dot\rho^2 )}\omega^2\dot\phi^2\right)^{1/2}.
	\label{eq:omega and r in loop}
\end{eqnarray}
The equation of motion for $\rho$ with the higher-dimensional winding (see eq.~\eqref{eq:v EOM2 r=omega}) 
can be represented in the form of a harmonic oscillator with the time-varying frequency:
\begin{eqnarray}
	\ddot\rho =-\frac{2\omega^2 -1}{\widetilde\epsilon^2}\rho .
\end{eqnarray}
The windings induce an additional, time-varying, energy injection to the four-dimensional part (although the overall energy is still conserved),
causing the loop to contract or expand at different rates.
Since $\rho$ and $\widetilde\epsilon$ are functions of $t$ only, due to the circular symmetry, 
this equation implies that the dependence on $\sigma$ in $\omega^2$ should vanish, namely
\begin{eqnarray}
	\partial_\sigma\omega^2 =0,
\end{eqnarray}
which is equivalent to the condition $(\widetilde\epsilon /\omega^2 )^{\cdot}=0$ 
from eq.~\eqref{eq:t EOM r=omega}.
Physically, this is because non-linear fluctuations in the winding density would induce
additional $\sigma$-dependence in the effective tension of the string, causing the loop to contract
or expand at different rates at different points on its circumference, in violation of our original assumption of circular symmetry.
One can also see that this has the stable loop solution $\dot\rho =0$ when $\omega^2 =1/2$, which
exactly coincides with the tensionless condition (see eq.~\eqref{eq:omega^2=1/2}), as expected. 
Moreover, since the condition \eqref{eq:condition} is given by
\begin{eqnarray}
	\widetilde\epsilon\,\dot\phi =\partial_\sigma\phi,
\end{eqnarray}
it is solved by the form
\begin{eqnarray}
	\phi =n_\theta\sigma +n_\theta\int^t\widetilde\epsilon^{-1}(t')dt'\ \ \ \ \ n_\theta \in \mathbb{Z},
\end{eqnarray}
where we have imposed the periodic boundary condition on $\phi$ and $n_\theta$ is the net number of winding
in the compact space (which are now distributed along the angular $\theta$-direction, $\theta \sim \sigma$).
Substituting this into eq.~\eqref{eq:omega and r in loop}, we obtain $\omega$ and $r$ : 
\begin{eqnarray}
	\omega^{-2}=r^{-2}=1+\left(\frac{Rn_\theta}{a\rho}\right)^2
	\,.
\end{eqnarray}
and the effective mass-density, shear and tension:
\begin{eqnarray} 
	\widetilde U=\mathcal{T}\Biggl[ 1+\left(\frac{Rn_\theta}{a\rho}\right)^2\Biggr]
	,\ \ 
	\widetilde\Sigma =\mathcal{T}\Biggl[ 1+\left(\frac{Rn_\theta}{a\rho}\right)^2\Biggr]^{-1}
	,\ \ 
	\widetilde T=\mathcal{T}\Biggl[ 1-\left(\frac{Rn_\theta}{a\rho}\right)^2\Biggr],
\end{eqnarray}
which are analogous to those obtained in eq.~\eqref{eq:U Sigma T in straight string} for long, straight
strings. In fact, the expressions for the components of the energy-momentum tensor become
equivalent to those for the long, straight string under the correspondence $\Delta /2\pi\leftrightarrow \rho$, $n_z \leftrightarrow n_{\theta}$. 
Thus, the tensionless condition for circular loops is also equivalent to eq.~\eqref{tensionless1} 
under this correspondence, so that eq.~\eqref{tensionless2} is equivalent to
\begin{eqnarray} \label{tensionless2}
	a\lambda_{\theta} =2\pi R
\end{eqnarray} 
for $\lambda_{\theta} = 2\pi\rho/n_{\theta} \leftrightarrow \lambda_z = \Delta/n_{z}$.

\subsection{Planar loops}
\label{sec:Non-circular loops}

Since the discussion in section \ref{sec:Strings with higher-dimensional windings}
does not depend on the gauge, we can consider a wide class of self-consistent solutions 
with higher-dimensional windings, even when we cannot apply the standard gauge condition in the large dimensions.
As a specific example, for a known planar, but non-circular loop configuration in $M^4$, we find solutions for the equations of motion
for the corresponding non-circular loop with higher-dimensional windings.
The functional form of the embedding in the Cartesian coordinate system is then
\begin{eqnarray} \label{Emb_6.2}
X^I=\left( t(\tau)=\zeta \tau, x(\tau ,\sigma )=\rho (\tau ,\sigma )\cos\sigma ,y(\tau ,\sigma )=\rho (\tau ,\sigma )\sin\sigma ,0,\phi (\tau ,\sigma )\right).
\end{eqnarray}
Instead of adopting the standard gauge in the large dimensions, we choose to identify the the space-like worldsheet coordinate 
with a fixed spatial coordinate in the background space.
Let us consider the four-dimensional components to construct the (albeit rather complicated) self-consistent equations of motion.
The length and the perpendicular velocity are given by
\begin{eqnarray} \label{dl_6.1}
	&&\frac{d\widetilde\ell}{d\sigma}=\sqrt{a^2\rho^2+a^2(\partial_{\sigma}\rho)^2}
	,\\
	&&\widetilde v_\perp^m
		=\frac{\rho\dot\rho}{\rho^2 +(\partial_\sigma\rho )^2}
			\Bigl(\rho\cos\sigma +\partial_\sigma\rho\sin\sigma\,,\,\rho\sin\sigma -\partial_\sigma\rho\cos\sigma\,,\, 0\Bigr),
\end{eqnarray}
which leads to
\begin{eqnarray}
	\frac{\widetilde v_{\perp}^2}{a^2} = \dot{\rho}^2\left(1 - \frac{(\partial_{\sigma}\rho)^2}{\rho^2+(\partial_{\sigma}\rho)^2}\right).
\end{eqnarray}
We then obtain the mass density in the large dimensions and the $\widetilde j$ parameter as
\begin{eqnarray}
	\widetilde\epsilon =\frac{\rho^2 +(\partial_\sigma\rho )^2}{\sqrt{\rho^2\left( 1-\dot\rho^2\right) +(\partial_\sigma\rho )^2}}
	,\ \ \ \ \ 
	\widetilde j=\frac{\dot\rho\partial_\sigma\rho}{\sqrt{\rho^2\left( 1-\dot\rho^2\right) +(\partial_\sigma\rho )^2}}.
\end{eqnarray}
Once we substitute these into eqs.~\eqref{eq:phi reduced EOM}-\eqref{eq:v EOM2 r=omega}, we have
the self-consistent set of the equations of motion for $\rho$ and $\phi$.
Formally, we may write down the specific expressions for the equations of motion, but
these cannot be evaluated without adopting a more specific ansatz for $\rho$.
However, it is sufficient for our purpose to see that a tensionless solution for the wound string exists.
Hence, we can conclude from the discussion in section \ref{sec:Strings with higher-dimensional windings}
that, without specifying the form of the radial coordinate $\rho(\tau,\sigma)$, the tensionless condition
always corresponds to a loop configuration for which $\omega^2 =r^2=1/2$, even for $\widetilde j\neq 0$.

\subsection{The general tensionless state} \label{The general tensionless state}

Before closing this section, we discuss the general conditions for tensionless strings.
Assuming the general tensionless condition $\omega^2 =r^2=1/2$, given in Sect. \ref{sec:Dynamics in warped geometries},
and substituting it into eqs.~\eqref{eq:t EOM r=omega}-\eqref{eq:v EOM2 r=omega}, the equations of motion are drastically simplified:
\begin{eqnarray}
\dot{\widetilde\epsilon} +\partial_\sigma\widetilde j=0
,\ \ \ 
\left(\widetilde v_\perp^2\right)^{\cdot} =0
,\ \ \ 
2\partial_\sigma\widetilde j\left( a^2-\widetilde v_\perp^2\right) -\left(\widetilde j-1\right)\partial_\sigma\left(\widetilde v_\perp^2\right) =0
.\label{eq:tensionless state}
\end{eqnarray}
Previously, we found that under the tensionless condition the energy conservation law from the four-dimensional point of view generally holds,
in addition to the vanishing of the acceleration felt by the string. To construct the nontrivial solution with vanishing tension, we need to fix the gauge condition.
Hence, we further assume $\widetilde j=0$, which corresponds to the standard gauge from the four-dimensional point of view. In this case, the equations above lead to
\begin{eqnarray}
\widetilde\epsilon =\widetilde\epsilon (\sigma )
,\ \ \ \ \ 
\widetilde v_\perp^2 ={\rm const}
.\label{eq:tensionless state}
\end{eqnarray}
With these conditions, we can solve the general equation for the extra-dimension, eq.~\eqref{eq:phi reduced EOM}:
\begin{eqnarray}
\phi =f\left( qt + q\int^\sigma_0\widetilde\epsilon (\sigma' ){\rm d}\sigma'\right)
\,,
\end{eqnarray}
where $f$ is an arbitrary function and $q$ is an arbitrary constant.

\section{Conclusion}
\label{sec:Conclusion}
%
\indent
To summarize, we have considered strings with arbitrary configurations in the Minkowski directions of a compactified spacetime, 
which also wrap and spin around an $S^1$ subcycle of constant radius.
We have developed a general method to evaluate the effective, four-dimensional energy-momentum tensor for the wound string, 
for any string configuration, and demonstrated the existence of a generic tensionless condition, in which the string remains static with respect to the large dimensions.
Specifically, we have demonstrated the equivalence of the Euler-Lagrange equation for the extra-dimensional embedding coordinate 
and the far simpler condition $\omega = r$, where $\omega$ represents the local fraction of the total string length, and $r$ represents the fraction of the local
perpendicular velocity (the only physical velocity of the string), in the Minkowski directions.
\\ \indent
Based on this formula, we have shown that a string with such a critical configuration, in which the tension vanishes locally at all points, {\it always} exists for $\omega^2 =r^2=1/2$.
In addition, we have shown that the self-consistent set of equations for the wound string are such that, for a given embedding corresponding to a known self-consistent ansatz in Minkowski space,
it is \emph{always} possible to construct a higher-dimensional ``generalization", for which the string retains its overall shape, with respect to the infinite dimensions, during its dynamical evolution. 
Physically, this occurs when the motion of the string in the compact space is  ``tuned" in such a way as to respect the symmetry of the four-dimensional embedding. This, in turn, is what allows 
the existence of a tensionless condition, in which this shape is preserved \emph{statically} from a four-dimensional perspective.
\\ \indent
Due to the formal correspondence between $M^4 \times S^1$ compactified strings and current-carrying strings, under dimensional reduction \cite{Ni79,NiOl87}, it follows that the phenomenon described in this 
analysis is analogous to that discovered previously for superconducting and chiral current-carrying strings. However, the models for which the existence of a generic tensionless state 
has (so far) been explicitly demonstrated are not completely general, and correspond to strings with neutral current, rather than to current-carrying strings with nonzero charge 
\cite{CoHiTu87,Test3,Test4,BlOlVi01}. 
The analysis of wound strings automatically avoids this restriction since the string momentum in the compact space may be reinterpreted as an electric charge, whereas as the corresponding 
momentum in the large directions (interpreted as the motion of the charge from a four-dimensional perspective) is separately conserved. 
As such, the formula presented here would be useful when one discusses a generalization, 
both of previous results for neutral current-carrying strings, and of the charged string vorton models 
\cite{Test1,Test2,Carter(1990),Carter&Martin(1993),Larsen(1993),Martins&Shellard(1998),Carter&Davis(2000),LaWa10,LaYo12,Vortons(2013)}, 
which correspond to the case of circular wound strings, here considered as a specific example. \\
\indent
Remarkably, from the formal correspondence demonstrated by the original, and extremely powerful, analysis by Nielsen \cite{Ni79}, it follows \emph{implicitly} that tensionless states exist generically for charged 
current-carrying strings, of arbitrary configuration, which therefore include all vorton-type species. 
As shown above, the higher-dimensional perspective, pioneered by Nielsen, proves to be extremely fruitful in considering such a general analysis and negates the need for the construction of an effective action.
\\ \indent
Finally, with regard to specific examples, we have explicitly constructed the self-consistent solution of the equations of motion 
for long, straight strings and circular loops with higher-dimensional windings, 
and have demonstrated that the tensionless condition corresponds to a configuration in which the wavelength of the windings, 
with respect the the four-dimensional space, is equal to their circumference (i.e. circumference of $S^1$ subcycle).
An important point to note is that the effective energy-momentum tensor for string with higher-dimensional
windings has, in general, a non-vanishing shear term, which may lead to distinguishable features in observations. 
Hence, the framework presented here may be useful in exploring ways to detect wound strings, and to distinguish them from other, current, charge and momentum carrying 
string species, whether these arise from field-theoretic or superstring models.

%
\begin{center}
{\bf Acknowledgments}
\end{center}
We would like to thank Mark Hindmarsh for useful discussions during the preparation of the manuscript.
D.Y. is supported by Grant-in-Aid for JSPS Fellows (No.259800).

%

 %
\end{document}